\documentclass[12pt]{article}
\usepackage{wrapfig}
\usepackage{epsfig}
\begin{document}
\begin{flushright}
JHU-TIPAC-020020\\
June 2002
\end{flushright}
\vspace{3mm}
\begin{center}
{\bf\large Low Scale String Unification and the Highest Energy Cosmic Rays}\footnote{Paper
submitted to ICHEP2002, Amsterdam}\\[3mm]
William S. Burgett$^a$,~G.~Domokos$^b$~and~S.~Kovesi-Domokos$^b$\\
$^a$ Department of Physics, University of Texas at Dallas;\\ e-mail: burgett@utdallas.edu\\
$^b$ Department of Physics and Astronomy, The Johns Hopkins University;\\ e-mail: skd@jhu.edu
\end{center} \vspace{3mm}
{\small {\em Abstract.}
String unification at a  scale of a few tens of TeV explains the existence of
cosmic ray interactions beyond the Greisen-Zatsepin-Kuzmin (GZK) cutoff.
Trans-GZK cosmic rays are neutrinos which can penetrate the cosmic microwave
background. In interactions with atmospheric nuclei they have sufficient energy
for exciting string modes. We present a model for the description of such
interactions and discuss the properties of the resulting extensive air showers.
Presently available data on trans-GZK cosmic rays suggest a string scale
around 80~TeV.}

\vspace{3mm}

It has been realized by Witten~\cite{witten1} some time ago that in certain 
strongly 
coupled string models, the multidimensional (d=10 or 11) string scale  and
the four dimensional Planck scale are less rigidly coupled than it was 
previously believed. This insight gave rise to a flurry of papers: several
authors pointed out that in such a scenario, the string scale could be of the order
 of a few TeV see~\cite{lykken, dimo}for the original papers. Hence, 
even experiments at the LHC could  provide some evidence for the existence
of extra dimensions and string excitations. Along the same lines, we 
pointed out that in such a scenario the problem of the trans-GZK cosmic ray interactions
may be resolved assuming that they are caused by high energy neutrinos~\cite{prl1}.
 In fact, a neutrino penetrates the cosmic microwave background radiation (CMBR)
 essentially uninhibited.
The CMS energy in a collision  with a CMBR photon is of the order of 100~MeV:
 the interaction mfp in the collision is essentially infinite. 
By contrast, in an interaction with a
nucleus in the atmosphere, the CMS energy is of the order of a  few hundred TeV: 
hence, string modes are  excited and the cross section  grows to a hadronic size.

Currently there is no string model known to  be  in agreement with experimental data. 
In particular, it is hard to calculate within the framework of a strongly coupled 
theory\footnote{By means of an explicit calculation, it was shown that weakly coupled
string models cannot explain the trans-GZK cosmic ray interactions,
 {\em cf}~\cite{cornet} and \cite{rattazzi}.}
For this reason, we abstract features of current models which are likely to be present
in future, phenomenologically successful theories.

The following basic ingredients are used:
\begin{itemize}
\item Unitarity of the $S$-matrix.
\item A  rapidly rising level density of resonances in dual models.
\item Unification of interactions at around the string scale, hereafter denoted by $M$.
\item Duality between resonances in a given channel and Regge exchanges in crossed channels.
\end{itemize}
Concerning the last item, it should be kept in mind that duality between resonances
and Regge {\em poles} is exact only in the tree approximation to a string amplitude.
It is unclear what the precise form of a generalization to world sheets of higher genus is:
probably, resonances of finite width are dual to Regge cuts. Thus, our formul{\ae}  are
likely to be valid to logarithmic accuracy.
Using these ingredients and the optical theorem, one obtains that the total cross section 
in a neutrino-parton interaction is\footnote{All energies are assumed to be large compared to
the rest energies of the incoming particles}:
\begin{equation}
\hat{\sigma} ( \hat{s} ) = \frac{8\pi}{\hat{s}}\sum_j^{N \left( \hat{s} \right)}
(2j + 1) \left( 1- \eta_j \cos \left( 2\delta_j \right)\right),
\label{partonsigma}
\end{equation}
where, as usual, $\eta$ and $\delta$ stand for the elasticity coefficient and phase shift
of a given partial wave, respectively. The quantity $N\left( \hat{s} \right)$ is the level of
the resonance, equal to the maximal angular momentum. 

For elastic resonances, $\eta = 1$ and $\delta \approx \pi /2$ within the width 
of the resonance. In that case, {\em on resonance} the total cross section is just proportional
to the number of states at a given level. Due to the finite widths of resonances in any
realistic model, it makes sense to average the cross section over an energy interval
comparable to the widths of the resonances. In such an approximation, one can introduce
the {\em density of states}, $d\left( \hat{s} \right)$ and regard $N$  a continuous variable,
such that $N \approx \hat{s} /M$\footnote{In the last formula, the Regge intercept
has been neglected. However, we shall see shortly that the excitations begin to
contribute significantly to the cross section for $N\geq 10$ or so; hence this approximation
is justified.}.
 Using this, one gets from eq.~(\ref{partonsigma}):
\begin{equation} 
\hat{\sigma} \approx \frac{16\pi}{\hat{s}} d\left(\hat{s} \right) 
\label{lowexcit}
\end{equation}
Clearly, as inelastic channels open up, the elasticity  coefficients in eq.~(\ref{partonsigma})
become less than unity and eq.~(\ref{lowexcit}) is no longer valid. Without any detailed 
knowledge of the inelastic channels (world sheets of a higher genus in present day 
string models), one can estimate the behavior of the cross section as 
$\hat{s} \rightarrow \infty$. Duality tells us that the leptoquark excitations should be dual
to the exchange of the $Z$-trajectory in the $t$-channel. 
Hence, apart from logarithmic corrections,
\begin{equation}
\hat{\sigma} \sim \hat{s}^{(\alpha (0) -1)},
\label{asympt}
\end{equation}
where $\alpha (0) $ is the intercept (branch point, respectively) of the $Z$ trajectory.
Apart from corrections of the order of $(M_{Z}/M)^{2}$, one has $\alpha (0) =1 $, so that
the neutrino-parton cross section tends to a constant. (We  verify {\em a posteriori}
that $M_{Z}/M \ll 1$, so that the power corrections to the cross section are insignificant
at all energies of interest.)

\begin{figure}[h]
\epsfig{file=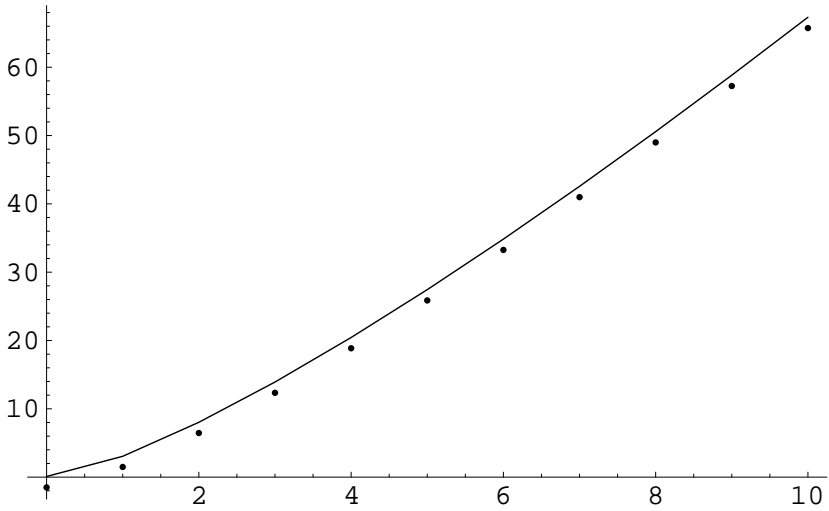, width=0.8\textwidth}
\label{leveldensity}
\caption{The level density of the RNS model. The points are from ref.~\cite{greenetal}; 
 the continuous curve is the best fit to the level density as explained in
 the text.}
\end{figure}

The level density is a rapidly rising function of $\hat{s} $. It is known that asymptotically
it rises as $\exp (a \sqrt{\hat{s}  /M})$, with $a$ being some constant;
 see, for instance~\cite{greenetal}.  
At the beginning of the spectrum, however, the rise is more rapid. 
The first few levels of the RNS model (\cite{greenetal} ({\em loc.cit})can be well interpolated by 
the function
\begin{equation}
d(N) \propto  \exp 1.24 N, \qquad N \approx \hat{s} /M ,
\label{density_of_states}
\end{equation}
see  Figure~\ref{leveldensity}. In this figure, the points have been calculated from the
generating function of the level density, \cite{greenetal}~eq.~(4.3.64).   
Other string models exhibit a similar rapid rise of the level
density.
Due to one's inability to carry out detailed calculations in a strongly coupled string model,
we chose to interpolate between the low excitation regime, eq~(\ref{lowexcit}) and the 
asymptotic one, eq.(\ref{asympt}. There are infinitely many functions, of course, 
interpolating between those limits: we were guided by a requirement of simplicity.
Having experimented with  a number of functional forms,  we came to the conclusion that,
after averaging over the parton distribution within the nucleon, the results are rather
insensitive to the precise form of the $\nu$-quark cross section. For that reason, we
chose a simple form satisfying the limits at low and high excitations:
\begin{equation}
\hat{\sigma} = \Theta \left(\hat{s}-M^{2}\right)  \frac{16\pi}{M^{2}}\frac{40 \exp 1.24 N_{0}}{1 +
 \frac{\hat{s}}{M^{2}}\exp 1.24 \left( N_{0} - \hat{s}/M^{2}\right)}
\label{interpolate}
\end{equation}

In eq.~(\ref{interpolate}), $M$ is the string scale and $N_{0}$ is a parameter measuring
the onset of the ``new physics''. In fact, one can convert that dimensionless parameter
into an energy scale. Using our previous relations, one can write $N_{0}\approx \hat{s_{0}}/M$,
or in terms of a laboratory energy of the incoming neutrino, $N_{0}\approx 2m\hat{E_{0}}/M$,
$m$ being the mass of the nucleon. In all these equations, the ``hat'' over the energies
 involved serves as a reminder that the quantities have to be integrated  over the parton
distribution. As usual, the conversion is carried out by means of  substitutions such as,
$\hat{s}= x s$ $x$ being the momentum fraction of a parton within the nucleon. The step
function is inserted because the cross section of the ``new physics'' vanishes at CM
energies below the mass of the first resonance.

Finally, the neutrino-nucleon cross section has to be constructed by integrating 
eq.~(\ref{interpolate}) over the parton distribution in the atmosphere. In order to do so, one
takes into account the fact that the dominant nuclei in the atmosphere (N, O) contain an
equal number of protons and neutrons. The parton distributions have been taken from
CTEQ6, \cite{cteq6}. The dominant contribution comes from valence quarks; gluons
do not contribute, since no presently known unification scheme contains ``leptogluons''.
Finally, the contribution of the sea is negligibly small, since the latter is concentrated
around $x=0$.

With the limited amount of information currently available on trans-GZK cosmic ray 
interactions, it is impossible to precisely determine the two parameters entering
eq.~(\ref{interpolate}). Nevertheless, the parameters can be bounded by the data.
From a qualitative point of view, the limitations come from the facts that
\begin{itemize}
\item No deep showers have been observed by Fly's Eye and HiRes.
\item The trans-GZK showers reported by AGASA, Fly's~Eye and Hi~Res appear to be
``hadron-like'', {\em i.e.} they originate high in the atmosphere and appear
to exhibit a development resembling proton induced showers.
\end{itemize}
Those constraints were analyzed by Sigl~{\em et al.} and Weiler,
 \cite{sigl, weiler}. In essence, the absence of deep showers excludes a region of the 
neutrino cross section, approximately, 0.02mb $\leq \sigma \leq $ 1mb. The cross section
has to grow to roughly hadronic size around the ``ankle'' in the cosmic ray spectrum,
 approximately at $5\times 10^{19}$eV
and stay of this size or grow slightly. 
Unless these conditions are satisfied, the neutrino
model of trans-GZK cosmic rays fails. 

\begin{figure}[h]
 \epsfig{file=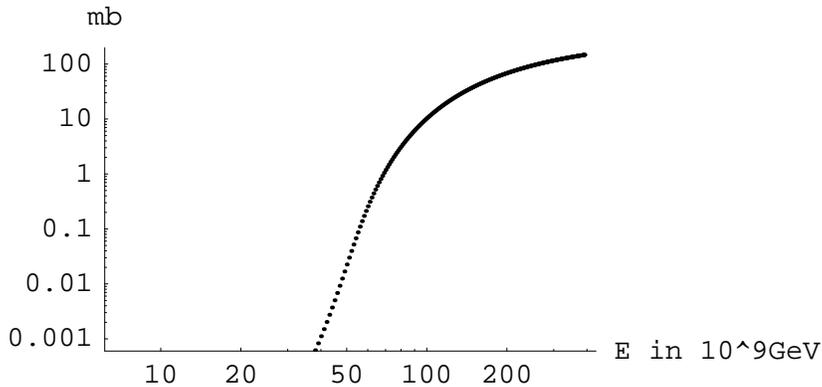, width=0.8\textwidth}
\label{crossection}
\caption{The $\nu$-nucleon cross section calculated from eq.~(\ref{interpolate}) and
the CTEQ6 parton distribution. $E_{0} = 5\times 10^{19}$eV, $M= 80$TeV.}
\end{figure}

A search of the parameter space yields reasonable
values for $E_{0}$ and $M$: $E_{0}\approx 5\times 10^{10}$GeV and $M\approx 80$TeV gives a 
cross section which is rising sufficiently rapidly: thus  it avoids the deep shower bound
and at the same time, it gives sufficiently large cross sections in the trans-GZK energy
region. These values of $E_{0}$ and $M$ give $N_{0}\approx 15.6$ confirming the intuitive
expectation. The neutrino-nucleon cross section with these values of the parameters is shown
in Fig.~(\ref{crossection}).
It is to be remarked that, due to the exponential dependence of eq.~(\ref{interpolate})
on the parameters, one cannot vary their values over a broad range without getting a
contradiction either with the bound on deep showers and/or with the required value
of the cross section for trans-GZK showers.

Neutrino induced showers were simulated using the ALPS (Adaptive Longitudinal Profile
Simulation) Monte Carlo package authored by Paul~T. Mikulski,\cite{pault}. Similarly to
earlier studies, see, {\em e.g.}~\cite{jhep} it was assumed that quarks and leptons are 
created in comparable numbers in an interaction as long as the CM energy of an interaction
 remains above 
$M$. Once the energy drops below $M$, the usual Standard Model cross sections
govern the further development of the shower. A qualitative consequence of this feature
is that, statistically, neutrino induced showers exhibit larger fluctuations than
proton induced ones, see~\cite{jhep}\footnote{This is a consequence of the fact that,
once in the standard model regime, leptonic interactions have a lower average multiplicity
than hadronic ones.}. Detailed results of such a simulation are
deferred to a forthcoming publication. Here we show an important characteristic
of the model: the depth of the initial interaction and its rms deviation.
\begin{figure}[h]
\epsfig{file=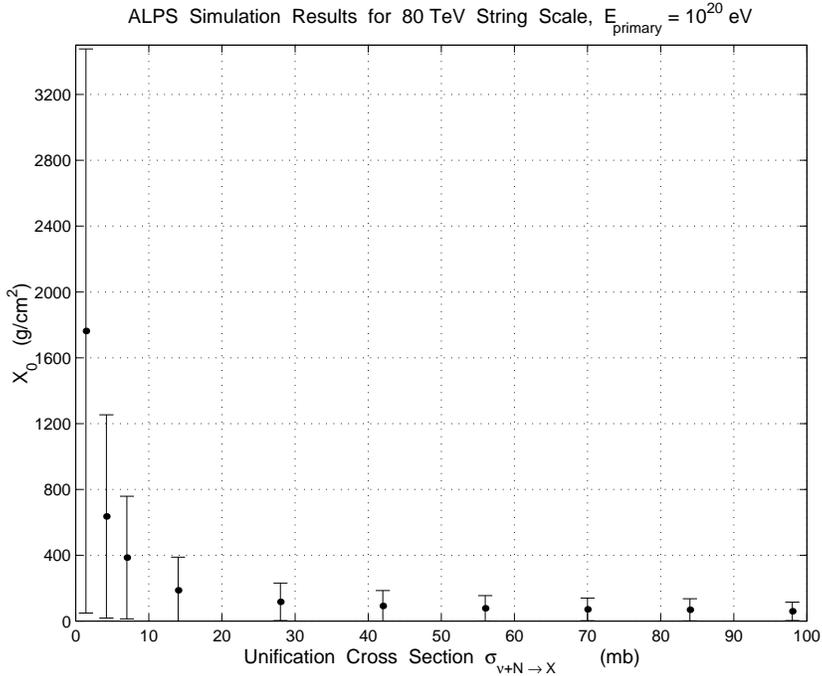, width=0.8\textwidth}
\label{initialinteraction}
\caption{Depth of the initial interaction of a neutrino within the framework of the 
unified model. The string scale is kept fixed at 80TeV; the cross ection is varied by
changing $E_{0}$.}
\end{figure}
It is clear from Fig.~\ref{initialinteraction} that once the cross section becomes larger than
about 20~mb or so, the shower starts high in the atmosphere. Hence, on an event by event basis, 
such showers are virtually indistinguishable from hadron induced showers. One will be able to
 test the validity of the scenario outlined here by a statistical study of the events observed
in future detectors, such as EUSO, OWL and the Pierre Auger Observatory.

\end{document}